\documentclass[prd,superscriptaddress,onecolumn,showkeys]{revtex4}
\usepackage{eurosym}
\usepackage{amsfonts}
\usepackage{array}
\usepackage{amsthm}
\usepackage{bm}
\usepackage{palatino}
\usepackage{mathpazo}
\usepackage{amssymb}
\usepackage{eurosym}
\usepackage{amsmath}
\usepackage{epsfig}
\usepackage{graphics}
\usepackage{changes}
\usepackage{color}
\usepackage{graphicx}
\usepackage[colorlinks=true,
            linkcolor=blue,
           urlcolor=blue,
           citecolor=blue]{hyperref}

\def\be{\begin{equation}}
\def\ee{\end{equation}}
\def\beq{\begin{eqnarray}}
\def\eeq{\end{eqnarray}}

\def\bes{\begin{eqnarray}}
\def\ees{\end{eqnarray}}

\begin{document}
\title{Tunneling of Massive Vector Particles from Types of BTZ-like Black Holes}

\author{Wajiha Javed}
\email{wajiha.javed@ue.edu.pk; wajihajaved84@yahoo.com} 
\affiliation{Division of Science and Technology, University of Education, Township-Lahore, Pakistan}

\author{Riasat Ali}
\email{riasatyasin@gmail.com}
\affiliation{Department of Mathematics, Government College,
University Faisalabad Layyah Campus, Layyah-31200, Pakistan}

\author{Rimsha Babar}
\email{rimsha.babar10@gmail.com} 
\affiliation{Division of Science and Technology, University of Education, Township-Lahore, Pakistan}

\author{Ali \"{O}vg\"{u}n}
\email{ali.ovgun@pucv.cl}
\homepage[]{https://www.aovgun.com}
\affiliation{Instituto de F\'{\i}sica, Pontificia Universidad Cat\'olica de Valpara\'{\i}%
so, Casilla 4950, Valpara\'{\i}so, Chile}
\affiliation{Physics Department, Faculty of Arts and Sciences, Eastern Mediterranean
University, Famagusta, North Cyprus, via Mersin 10, Turkey}

\begin{abstract}
In this paper, we analyze the Hawking radiation phenomenon for types
of Banados-Teitelboim-Zanelli-like (BTZ-like) black holes. For this purpose,
using the Hamilton-Jacobi method, we consider semi-classical WKB
approximation to calculate the tunneling probabilities of massive boson
particles. For these particles, we use the equation of motion for the Glashow-Weinberg-Salam model. Using
quantum tunneling process of charged massive bossons, we compute the corresponding
Hawking temperatures. Furthermore, we discuss the effects of rotation parameter on
tunneling probability and temperature.
\end{abstract}

\date{\today}
\maketitle
\section{Introduction}

The hypothesis of general relativity (GR) suggested that a sufficiently compact
mass can change the shape of spacetime fabric for configuration of a black hole (BH).
Hawking established that quantum events reserved BH to emit radiation. These radiation 
emit from a BH are just like a black body radiation and
known as Hawking radiation \cite{C5}. The Hawking radiation process is
studied by creation of a couple of positive and negative charged energy
particles in the vicinity of the horizon of BH. The positive-energy particle
escape out from horizon, causes Hawking radiation while the negative-energy
particle of this pair falls inside the BH. The rate at which particles overcome
the horizon can be calculated by quantum tunneling phenomenon. According
to this technique particles have finite tunneling probability
denoted by $\tilde{\Gamma}$ to cross the event horizon, which depends
upon the imaginary part of particle's action \cite{R1,R2}. The
probability $\tilde{\Gamma}$ for emitted particles can be defined as
\begin{equation}
\tilde{\Gamma} \varpropto \exp\left[-\frac{2}{\hbar}Im\pounds\right],\label{W1}
\end{equation}
where $\pounds$ denotes the semi-classical particle's action and $\hbar$
denotes the Planck's constant.

The study of Hawking temperature by utilizing quantum tunneling
process from various BHs has attracted lots of
researchers. Different efforts have been made to calculate this
radiation spectrum by considering quantum mechanics of
fermions, bosons, scalars, Dirac and photon particles etc. Various authors
\cite{R5,Sakalli:2015taa,Sakalli:2014sea,Sakalli:2016mnk,Sakalli:2017ewb,Ovgun:2015box,Jusufi:2015mii,Sakalli:2015mka,Sakalli:2015nza,Sakalli:2016cbo,Ovgun:2015jna,Ovgun:2017hje,Jusufi:2017vhz,Kuang:2017sqa,Kuang:2018goo,Javed:2018msn,Gonzalez:2017zdz,Ovgun:2016roz,Jusufi:2016hcn,deyou1,deyou2,aa19,aa20,aa21,za4,za5,za6,za7,za2,za3,R6,R7,R8} analyzed vector particles, fermionic particles,  spin-2 and spin-3/2 particles tunneling to
study the Hawking radiation phenomenon for different BHs and wormholes.
The tunneling probabilities for Kerr-Newman BH \cite{R3} and
charged black string \cite{R4} have also been investigated.
In order to study Hawking temperature by using Kerner and Mann's formulation, Sharif
and Javed \cite{a4} studied the fermions tunneling phenomenon
through the horizons of charged anti-de Sitter BHs, charged torus-like BHs,
$Pleba\acute{n}ski-Demia\acute{n}ski$ family of BHs, regular BHs and
and traversable wormholes. The same authors \cite{C1} discussed
the Hawking radiation phenomenon as fermions tunneling for a pair of
charged accelerating and rotating BHs with NUT parameter. They
have calculated the corresponding Hawking temperature. Sharif and Javed
\cite{A7} have also investigated the quantum corrections for regular
BHs, i.e., Bardeen and ABGB. Javed, Abbas and Ali
discussed charged vector particles tunneling process for a pair of accelerating
and rotating BHs as well as for 5D gauged super-gravity BHs \cite{[6]}. The
Hawking radiation for charged NUT (Newman-Unti-Tamburino)
BH having rotation and acceleration parameters has been viewed in Ref.\cite{R11}.

The tunneling rate for outgoing particles is found by utilizing the
imaginary component of particle's action. $\ddot{\textmd{O}}$vg$\ddot{\textmd{u}}$n,
Javed and Ali \cite{B56} found the tunneling rate of charged massive bosons for various
types of BHs surrounded by the perfect fluid in Rastall theory.
Several efforts \cite{R31}-\cite{R35} have been made to analyze the tunneling
process of charged and uncharged scalar and Dirac particles for different BHs.
The tunneling rate of spin-$\frac{1}{2}$ particles through horizon of Rindler
spacetime was discussed as well as the Unruh temperature was analyzed
\cite{R12}. Kraus and Wilczek \cite{R14, R15} visualized the
semi-classical method to study the Hawking radiation as a tunneling
through the horizon. This method contains the calculation for the process of
s-wave emission across the horizon.

In continuation of previous work, in this paper we calculate the electromagnetic
boson particles radiation, emitted by BTZ-like BHs. For this analysis, we study the
tunneling of massive vector bosons through event horizons of higher dimensional
charged BTZ-like BHs by using semi-classical approximation, also we
calculate the corresponding Hawking temperatures. Charged vector
particles (spin-1) such as charged $W^{\pm}$-bosons having much
importance in Standard Model. In the background of BHs
geometries; the behavior of the bosons particles can
be traced out by using Proca equation. Here, we analyze the field
equations of charged $W^{\pm}$-bosons by applying the Lagrangian of
\textmd{Glashow}-\textmd{Weinberg}-\textmd{Salam} model \cite{R21}.

This paper is organized as follows: In \textbf{Section} \textbf{II}, we analyze
the tunneling probability $\tilde{\Gamma}$ of boson particles and the
corresponding Hawking temperature $T_{H}$ for charged BTZ magnetic
static BH, by calculating the radial part of particle's action for
massive bosons $W^{\pm}$. \textbf{Section} \textbf{III} presents the vector
particles tunneling for rotating BTZ magnetic BH. \textbf{Section} \textbf{IV}
provides the results of $\tilde{\Gamma}$ and $T_{H}$ for $(n+1)$-dimensional BTZ
magnetic BHs with rotation parameter. \textbf{Section} \textbf{V} contains the
results of tunneling probability and Hawking temperature for
$(n+1)$-dimensional BTZ magnetic BHs with more than one rotation parameters.
\textbf{Section} \textbf{VI} summarizes all the results of the paper.

\section{Charged BTZ Magnetic Static Black Hole}

Black holes play a very essential part in the field of General Relativity.
After the discovery of first Schwarzschild BH many extensions have been made
by introducing the rotation, charge, acceleration, cosmological,
quintessential and many other parameters. All these extensions of
BHs are associated with D-branes. The BTZ solution in $(2 + 1)-$dimensional
space presents a valuable template to guide people in order to analyze concepts
of quantum gravitational problems and AdS/CFT conjecture. In a charged
BH the electric field is related to the temporal factor $A_{t}$
of gauged potential whereas the magnetic field is related to the angular component $A_{\phi}$.
As a result, one can expect the magnetic solution to be written in a
metric gauge, in which the $g_{tt}$ and $g_{\phi\phi}$ components are
relatively exchanged for their effects
to those used in electric gauge to describe BH solutions \cite{r1}.
The motivation of our work is that we are looking for a charged BH with magnetic field
instead of BH interpretation with the electrical field.

The background geometry of BTZ BH can be studied for more generalized, i.e.,
$n$ and $(n+1)-$dimensional spaces and it can be used to explain the effects
of more rotation parameters involve in BH to investigate either they will
effect or not the tunneling spectrum of Hawking
radiation. There have been a lot of work on quantum
tunneling spectrum for charged BTZ BH but this form of BTZ-like BH
is more generalized than the BTZ BH. The line-element of this BH is given by \cite{24}
\begin{equation}
ds^{2}=-\frac{r^{2}}{l^{2}}dt^{2}+\frac{1}{R(r)}dr^{2}
+l^{2}R(r)\gamma^2 d\phi^{2},\label{111}
\end{equation}
where $r$ is radial coordinate, $l$ denotes the curvature radius associated with negative
cosmological constant $\Lambda=-\frac{1}{l^{2}}$, $R(r)$ represents the metric function which is given by
\begin{equation}
R(r)=\frac{r^{2}}{l^{2}}-\left[M+8q^{2}l^{2}\gamma^{2}
 \ln\left(\frac{r}{l}\right)\right],
\end{equation}
where $M$ and $q$ indicate the mass and charge of BTZ solution,
while $\gamma$ is arbitrary parameter.

It is to be noted that the above metric (\ref{111}) after applying local transformation
$t\rightarrow il\gamma \phi$ and $\phi\rightarrow it/l$
will convert to the static 3D Schwarzschild metric. So, we can change the role of
$t$ and $\phi$ co-ordinates by using given transformation.
The Eq.(\ref{111}) can be rewritten as
\begin{equation}
ds^{2}=X(r)dt^{2}+Y(r)dr^{2}
+Z(r)d\phi^{2},
\end{equation}
where the radial functions $X(r)$, $Y(r)$ and $Z(r)$ are expressed as
\begin{equation}
X(r)=-\frac{r^{2}}{l^{2}},~~
Y(r)=\frac{1}{R(r)},~~
Z(r)=l^{2}R(r)\gamma^2.
 \end{equation}
The gauge potential of this BH can be defined as
\begin{equation}
A_{\mu}=-2ql^{2} \gamma^{2} h(r)\delta^{\phi}_{\mu},
\end{equation}
where the only non-zero component is $A_{\phi}$ due to magnetic
field and $h(r)$ is an arbitrary function of $r$.

In order to calculate the tunneling rate for massive vector
particles through the BH's event horizon $r_+$, we will consider the
relativistic wave equation, i.e., Proca equation with electromagnetic effects. The motion of charged massive
spin-$1$ fields is traced out by the given Proca equation in the background of electromagnetic field $\Psi_{\mu}$ \cite{[6]}
\begin{equation}
\frac{1}{\sqrt{-\textbf{g}}}\partial_{\mu}(\sqrt{-\textbf{g}}\Psi^{\nu\mu})+
\frac{m^{2}}{h^{2}}\Psi^{\nu}+\frac{\iota}{h}e A_{\mu}\Psi^{\nu\mu}+
\frac{\iota}{h}eF^{\nu\mu}\Psi_{\mu}=0,\label{21}
\end{equation}
here $g$ indicates the determinant of the coefficient matrix, $e$ is
particle charge and $\Psi^{\mu\nu}$ is anti-symmetric tensor which can be defined as
\begin{equation}
\Psi_{\nu\mu}=\partial_{\nu}\Psi_{\mu}-\partial_{\mu}\Psi_{\nu}+
\frac{\iota}{h}e A_{\nu}\Psi_{\mu}-\frac{\iota}{h}e A_{\mu}\Psi_{\nu}~~
\textmd{and}~~
F^{\mu\nu}=d^{\mu}A^{\nu}-d^{\nu}A^{\mu},
\end{equation}
where, $A_{\mu}$ is the electromagnetic vector potential related
to BH and $d^{\mu}$ refers to the geometrical
covariant derivative, $m$ represents the particle's mass.
The non-zero components of $\Psi^{\mu}$ and
$\Psi^{\nu\mu}$ are given as follows
\begin{eqnarray}
\Psi^{0}&=&\frac{\Psi_{0}}{X(r)},~~~~~~~~~ \Psi^{1}=
\frac{\Psi_{1}}{Y(r)},~~~~~~~~~~ \Psi^{2}
=\frac{\Psi_{2}}{Z(r)},\\
\Psi^{01}&=&\frac{\Psi_{01}}{X(r)Y(r)},~~~
\Psi^{02}=\frac{\Psi_{02}}{X(r)Z(r)},~~~
\Psi^{12}=\frac{\Psi_{12}}{Y(r)Z(r)}.
\end{eqnarray}
Using Hamilton-Jacobi ansatz, the wave function in terms of semi-classical particle's action $\pounds$ by
applying the WKB approximation can be defined as \cite{[6],R11}, \cite{2}-\cite{[7]}
\begin{equation}
\Psi_{\nu}=c_{\nu}\exp\left[\frac{\iota}{\hbar}\pounds_{0}(t,r,\phi)+
\sum_{i=1}^{n} \hbar^{i}\pounds_{i}(t,r,\phi)\right].\label{rias}
\end{equation}
Using Eq.(\ref{rias}) in the wave Eq.(\ref{21}) for $i=1,2,3,...$
and by neglecting the higher order
terms, we get the following set of equations
\begin{eqnarray}
&&\frac{1}{Y(r)}[c_{1}(\partial_{0}\pounds_{0})
(\partial_{1}\pounds_{0})-c_{0}(\partial_{1}\pounds_{0})^{2}]+\frac{1}{Z(r)}
[c_{2}(\partial_{0}\pounds_{0})(\partial_{2}\pounds_{0})
-c_{0}(\partial_{2}\pounds_{0})^{2}\nonumber\\
&&-eA_{\phi}c_{0}(\partial_{2}\pounds_{0})]+\frac{eA_{\phi}}{Z(r)}[c_{2}
(\partial_{0}\pounds_{0})-eA_{\phi}c_{0}-c_{0}(\partial_{2}\pounds_{0})]
-m^{2}c_{0}=0,\label{31}\\
&&\frac{1}{X(r)}[c_{0}(\partial_{0}\pounds_{0})
(\partial_{1}\pounds_{0})-c_{1}(\partial_{0}\pounds_{0})^{2}]+
\frac{1}{Z(r)}[c_{2}(\partial_{1}\pounds_{0})
(\partial_{2}\pounds_{0})-c_{1}(\partial_{2}\pounds_{0})^{2}\nonumber\\
&&-eA_{\phi}c_{1}(\partial_{2}\pounds_{0})]+\frac{eA_{\phi}}{Z(r)}
[c_{2}((\partial_{1}\pounds_{0})-eA_{\phi})c_{1}-c_{1}(\partial_{2}\pounds_{0})]
-m^{2}c_{1}=0,~~~~~\label{41}\\
&&\frac{1}{X(r)}\left[c_{0}(\partial_{0}\pounds_{0})
(\partial_{2}\pounds_{0})-c_{2}(\partial_{0}\pounds_{0})^{2}
+eA_{\phi}(\partial_{0}\pounds_{0})c_{0}\right]
+\frac{1}{Y(r)}[c_{1}(\partial_{1}\pounds_{0})\nonumber\\
&&(\partial_{2}\pounds_{0})
-c_{2}(\partial_{1}\pounds_{0})^{2}+eA_{\phi}c_{1}
(\partial_{1}\pounds_{0})]-m^{2}c_{2}=0.\label{51}
\end{eqnarray}
Using separation of variables technique, one can write the particle's action in the following form
\begin{equation}
\pounds_{0}=-(E-j\omega)t+G(r)+k\phi,
\end{equation}
where $E$, $\omega$ and $j$ represent particle's energy, angular velocity and angular momentum,
respectively. Using Eqs.(\ref{31})-(\ref{51}), the matrix equation can be written as
\begin{equation}
\Xi(c_{0},c_{1},c_{2})^{T}=0,
\end{equation}
where $\Xi$ is a $3\times3$ ordered matrix and its components are given below
\begin{eqnarray}
\Xi_{00}&=&-\frac{\acute{G}^{2}}{Y(r)}-\frac{1}{Z(r)}(k^{2}
-eA_{\phi}k)-m^{2}-\frac{eA_{\phi}}{Z(r)}(k+eA_{\phi}),\\
\Xi_{01}&=&-\frac{1}{Y(r)}\acute{G}(E-j\omega),\\
\Xi_{02}&=&-\frac{1}{Z(r)}(E-j\omega)k-\frac{eA_{\phi}}{Z(r)}(E-j\omega),\\
\Xi_{10}&=&-\frac{1}{X(r)}(E-j\omega)\acute{G},\\
\Xi_{11}&=&-\frac{1}{X(r)}(E-j\omega)^{2}-\frac{1}{Z(r)}
(k^{2}+eA_{\phi}k)-m^{2}-\frac{eA_{\phi}}{Z(r)}(k+eA_{\phi}),\\
\Xi_{12}&=&\frac{1}{Z(r)}\acute{G}k+\frac{eA_{\phi}}{Z(r)}\acute{G},\\
\Xi_{20}&=&-\frac{1}{X(r)}\left[(E-j\omega)k+eA_{\phi}(E-j\omega)\right],\\
\Xi_{21}&=&\frac{1}{Y(r)}[\acute{G}k+eA_{\phi}\acute{G}],\\
\Xi_{22}&=&-\frac{1}{X(r)}(E-j\omega)^{2}
%\deleted{+eA_{\phi}(E-j\omega)}
-\frac{1}{Y(r)}\acute{G}^{2}-m^{2},
\end{eqnarray}
where $\acute{G}=\partial_{r}\pounds_{0}$ and
$k=\partial_{\phi}\pounds_{0}$.
For the non-trivial solution, we take $\mid\Xi\mid=0$ and get
\begin{equation}
\acute{G}(r)=\sqrt{\frac{\left(E-eA_{\phi}-j\omega\right)^{2}+X}
{\frac{-Z(r)}{Y(r)}}},\label{RA}
\end{equation}
where
\begin{equation}
X=(E-j\omega)k+k^{2}+m^{2}-eA_{\phi}k.
\end{equation}
Integrating around the pole at the horizon, Eq.(\ref{RA}) provides
\begin{equation}
ImG^{\pm}(r)=\pm\iota\pi\frac{(E-j\omega-eA_{\phi})}{2\tilde{\kappa}_{(r_{+})}},
\end{equation}
where $G^{+}$ and $G^{-}$ represent radial function for the outgoing and
incoming particles, respectively following the radial path. Therefore, particle's action $\pounds$ is reduce to its radial part $G$.
The surface gravity $\tilde{\kappa}(r_{+})$
of this BH is given as follows
\begin{equation}
\tilde{\kappa}(r_{+})=\frac{r_{+}^{2}-4q^{2}\gamma^{2}l^{4}}{l^{2}r_+}.
\end{equation}
Using Eq.(\ref{W1}), the tunneling probability $\tilde{\Gamma}$ for boson particles can be defined as ratio of probabilities of emission and absorption, i.e.,
\begin{eqnarray}
\tilde{\Gamma}&=&\frac{\tilde{\Gamma}_{(emission)}}{\tilde{\Gamma}_{(absorption)}}=
\frac{\exp\left[-\frac{2}{\hbar}(ImG^{+})\right]}{\exp\left[-\frac{2}{\hbar}(ImG^{-})\right]}=
\exp\left[-\frac{4}{\hbar}ImG^{+}\right]\nonumber\\
&=&\exp\left[-2\pi\frac{l^{2}(E-j\omega-eA_{\phi})}
{r_{+}^{2}-4q^{2}\gamma^{2}l^{4}}\right]\quad\quad \textmd{(for $\hbar=1$)}.
\end{eqnarray}
Using Boltzmann factor $\tilde{\Gamma}_{B}=
\exp\left[(E-j\omega-eA_{\phi})/T_{H}\right]$,
the corresponding Hawking temperature can be calculated as
\begin{equation}\label{aTH}
\tilde{T}_{H}=\frac{\tilde{\kappa}}{2\pi}=\frac{r_{+}^{2}-4q^{2}\gamma^{2}l^{4}}{2\pi l^{2}r_+}.
\end{equation}
The tunneling probability of boson particles depending upon
charged potential of BH, particle's charge, energy and angular momentum,
the curvature radius $\ell$
and BH radius. While,
the Hawking temperature is depending upon
charge of BH, the curvature radius and BH radius.

\section{Rotating BTZ Magnetic Black Hole}

In this section, we analyze the metric by using co-ordinate transformation such as
$t\longrightarrow\lambda t-a\varphi$ and
$\varphi\longrightarrow\lambda\varphi-\frac{a}{l}t.$
One can obtain the BTZ BH with rotation parameter as well as magnetic charge.
The line-element of this BH is given by \cite{24}
\begin{eqnarray}
ds^{2}&=&-\frac{r^{2}}{l^{2}}(\lambda dt-ad\varphi)^{2}+\frac{1}{R(r)}dr^{2}
+l^{2}R(r)\gamma^{2} (\frac{a}{l^2}dt-\lambda d\varphi)^{2},\label{1}
\end{eqnarray}
where $\lambda=\sqrt{1+\frac{a^2}{l^2}}$ and $a$ is a rotation parameter.
The gauge potential $A_{\mu}$ for this BH is given by
\begin{equation}
A_{\mu}=-2ql\gamma^{2} h(r)\left(\frac{a}
{l}\delta^{t}_{\mu}-\lambda l\delta_{\mu}^{\varphi}\right).
\end{equation}
The non-zero components are $A_{t}$ and $A_{\varphi}$.

Following the procedure given in the preceding \textbf{Section} \textbf{II}, we can
obtain the surface gravity $\tilde{\kappa}(r_{+})$ of rotating
BTZ-like magnetic BH in the following form
\begin{equation}
\tilde{\kappa}(r_{+})=\frac{\lambda^{2} r_{+}^{2}}{l^{3}}
-\frac{r_{+}^{2}\gamma^{2}a^{2}}{l^{2}}+4q^{2}\gamma^{4}r_{+}a^{2}
-\frac{4\lambda^{2} q^{2}\gamma^{2}}{l^2}+
4q^{2}l^2 \gamma^{4}a^{2}-\frac{16l^{2}q^{4}\gamma^{6}a^{2}}{r_{+}}.
\end{equation}
The tunneling probability $\tilde{\Gamma}$ for this case of BTZ BH can be obtained as
\begin{equation}
\tilde{\Gamma}=\exp\left[\frac{-2\pi(E-j\omega-eA_{t}-eA_{\varphi})}
{\frac{\lambda^{2} r_{+}^{2}}{l^{3}}
-\frac{r_{+}^{2}\gamma^{2}a^{2}}{l^{2}}+4q^{2}\gamma^{4}r_{+}a^{2}
-\frac{4\lambda^{2} q^{2}\gamma^{2}}{l}+
4q^{2}l \gamma^{4}a^{2}-\frac{16l^{2}q^{4}\gamma^{6}a^{2}}{r_{+}}}\right].
\end{equation}
Using Boltzmann factor, we can obtained the corresponding
Hawking temperature $\tilde{T}_{H}$ for boson particles as
\begin{equation}\label{aTH}
\tilde{T}_{H}=\frac{1}{\pi^{2}}\left[\frac{\lambda^{2} r_{+}^{2}}{4l^{3}}
-\frac{r_{+}^{2}\gamma^{2}a^{2}}{4l^{2}}+q^{2}\gamma^{4}r_{+}a^{2}
-\frac{\lambda^{2} q^{2}\gamma^{2}}{l^2}+
q^{2}l^2 \gamma^{4}a^{2}-\frac{4l^{2}q^{4}\gamma^{6}a^{2}}{r_{+}}\right].
\end{equation}
Thus, the surface gravity $\tilde{\kappa}(r_{+})$, tunneling rate $\tilde{\Gamma}$
and Hawking temperature $\tilde{T}_{H}$, all are depending upon rotation parameter.

\section{$(n+1)$-dimensional Charged BTZ Black Hole}

In this section, we calculate the massive boson particles tunneling probability
at the outer horizon of the $(n+1)$-dimensional BTZ magnetic BH with
rotation parameter. The line-element of this BH is given by \cite{24}
\begin{eqnarray}\nonumber
ds^{2}&=&-\frac{r^{2}}{l^{2}}(\lambda dt-ad\varphi)^{2}+\frac{1}{R(r)}dr^{2}
+l^{2}R(r)\gamma^{2} (\frac{a}{l^{2}}dt-\lambda d\varphi)^{2}+r^{2}d\phi^{2}\nonumber\\
&+&\frac{r^{2}}{l^{2}}\sum_{i=1}^{n-3}dx^{i}.\label{1}
\end{eqnarray}
It is to be noted that the $x^{i}$ is the dimension length of BH and the angular
parameter $\varphi$ and $\phi$ are dimensionless for $[0,2\pi]$.
For this case, the metric function $R(r)$ is defined as
\begin{equation}
R(r)=\frac{r^{2}}{l^{2}}-\frac{1}{r^{d-3}}
\left[M+2^{(d-1)/2}(2ql\gamma)^{d-1}
\ln\left(\frac{r}{l}\right)\right].
\end{equation}
Using the same formalism as defined earlier in preceding \textbf{Section} \textbf{II} for this line-element,
the surface gravity $\tilde{\kappa}(r_{+})$ can be obtained as
\begin{equation}
\tilde{\kappa}(r_{+})=\frac{\lambda^{2}\acute{R}(r_{+}) r_{+}}
{2l^{2}}+\frac{\acute{R}^{2}(r_{+})a^{2}\gamma^{2}}{4},
\end{equation}
where $\acute{R}(r)$ is defined as
\begin{eqnarray}
\acute{R}(r_{+})&=&\frac{2r_{+}}{l^{2}}-\frac{M(3-d)}{r_{+}^{d-2}}
+\frac{(3-d)}{r_{+}^{d-2}}2^{(d-1)/2}(2ql\gamma)^{d-1}
\ln\left(\frac{r_{+}}{l}\right)\nonumber\\
&+&\frac{2^{(d-1)/2}(2ql\gamma)^{d-1}}{r_{+}^{d-2}}.
\end{eqnarray}
The tunneling rate $\tilde{\Gamma}$ of boson particles
at horizon can be calculated as follows
\begin{equation}
\tilde{\Gamma}=\exp\left[\frac{-2\pi l^2(E-j\omega-eA_{\mu})}
{2\lambda^{2}\acute{R}(r_{+}) r_{+}
+l^{2}\acute{R}^{2}(r_{+})a^{2}\gamma^{2}}\right].
\end{equation}
The corresponding Hawking temperature $\tilde{T}_{H}$ is given by
\begin{equation}\label{aTH}
\tilde{T}_{H}=\frac{2\lambda^{2}\acute{R}(r_{+})r_{+}+
\acute{R}^{2}(r_{+})l^{2}a^{2}\gamma^{2}}{4\pi l^2}.
\end{equation}
The tunneling rate $\tilde{\Gamma}$ and Hawking temperature
$\tilde{T}_{H}$ of a boson particles depend on rotation parameter, mass and BH radius.

\section{$(n+1)$-dimensional Charged BTZ Black Hole with $[n/2]$ Rotation Parameters}

This section is devoted to analyze the massive boson particles tunneling probability
at the outer horizon of the metric of $(n+1)$-dimension BH
with more rotation parameters and charge. The line-element of this BH is given by \cite{24}
\begin{eqnarray}\nonumber
ds^{2}&=&-\frac{r^{2}}{l^{2}}\left[\lambda
 dt-\sum_{i=1}^{k} a_{i}d\varphi^{i}\right]^{2}+\frac{1}{R(r)}dr^{2}
+R(r)\gamma^{2}\left[\sqrt{\lambda^{2}-1}dt-
\frac{\lambda}{\sqrt{\lambda^{2}-1}}\right.\nonumber\\
&&\left.\sum_{i=1}^{k} a_{i}d\varphi^{i}\right]^{2}
+\frac{r^{2}}{l^{2}(\lambda^{2}-1)}\sum_{i<j}^{k}
\left[a_{i}d\varphi_{j}-a_{j}d\varphi_{i}\right]^{2}
+r^2d\varphi^{2}+\frac{r^{2}}{l^{2}}dX^{2}.\label{Mr}
\end{eqnarray}
There are $[n/2]$ independent rotation parameters.
It is to be noted that $\lambda=\sqrt{1+\sum_{i=1}^{k}\frac{a_{i}^{2}}{l^2}}$
and $dX^{2}$ is the Euclidean metric
on the $(n-k-2)$-dimensional subspace. The function $R(r)$ is defined as
\begin{equation}
R(r)=\frac{r^{2}}{l^{2}}-\frac{1}{r^{d-3}}
\left[M+2^{(d-1)/2}(2ql\gamma)^{d-1}
\ln\left(\frac{r}{l}\right)\right].
\end{equation}
Following the same procedure given in the preceding sections, the surface gravity $\tilde\kappa(r_{+})$ of
$(n+1)$-dimensional BTZ-like BH with more rotation parameter can be calculated as
\begin{equation}
\tilde{\kappa}(r_{+})=\frac{\lambda^{2}\acute{R}(r_{+}) r_{+}}
{2l^{2}}+\frac{\acute{R}^{2}(r_{+})(\lambda^{2}-1)\gamma^{2}}{4}.
\end{equation}
The tunneling rate $\tilde{\Gamma}$ of boson particles at
event horizon can be calculated as
\begin{equation}
\tilde{\Gamma}=\exp\left[\frac{-2\pi l^2(E-j\omega-eA_{\mu})}
{2\lambda^{2}\acute{R}(r_{+}) r_{+}
+l^{2}\acute{R}^{2}(r_{+})(\lambda^{2}-1)\gamma^{2}}\right].
\end{equation}
The corresponding Hawking temperature $\tilde{T}_{H}$ can be deduced as
\begin{equation}\label{aTH}
\tilde{T}_{H}=\frac{2\lambda^{2}\acute{R}(r_{+}) r_{+}+
\acute{R}^{2}(r_{+})l^{2}(\lambda^{2}-1)\gamma^{2}}{4\pi l^2},
\end{equation}
where
\begin{eqnarray}
\acute{R}(r_{+})&=&\frac{2r_{+}}{l^{2}}-\frac{M(3-d)}{r_{+}^{d-2}}
+\frac{(3-d)}{r_{+}^{d-2}}2^{(d-1)/2}(2ql\gamma)^{d-1}
\ln\left(\frac{r_{+}}{l}\right)\nonumber\\
&+&\frac{2^{(d-1)/2}(2ql\gamma)^{d-1}}{r_{+}^{d-2}}.
\end{eqnarray}
It is to be noted that $\lambda=\sqrt{1+\sum_{i=1}^{k}\frac{a_{i}^{2}}{l^2}}$ depends
upon more rotation parameters. Hence, the surface gravity, tunneling probability and
the Hawking temperature for vector particles are depending
upon more rotation parameters.
The Hawking temperature increases due to rotation parameter.

\section{Conclusions}

In this work, we have studied the massive boson
particles tunneling phenomenon from 3-dimensional static charged BTZ magnetic
as well as rotating BH and $(n+1)$-dimensional charged BTZ BH
with rotation parameter. We extended our analysis
to higher-dimensional rotating charged BTZ-like BHs. For this purpose,
initially we have utilized the equation of motion for massive bosons by generalizing
the Proca wave equation of charged particles in curved spacetime in the
background of electromagnetic field.
We have applied the WKB approximation to the
Proca equation and in result we have obtained a set of field equations, then
we have used the separation of variables technique to solve these equations. We have obtained the radial function of semi-classical particle's action by putting the determinant of coefficient matrix
equals to zero. By utilizing the obtained surface gravity,
we have calculated the tunneling probability and Hawking temperature
for all types of given BTZ-like BHs.

The tunneling probabilities and Hawking temperatures are depending upon the parameters on which BHs are dependent.
It is interesting to specify here that the back-reaction impacts of the radiated
particles on the BH geometry and self-gravitating effects have been
ignored and the determined Hawking temperature
is just a main term and turned out to be reliable with BH universality.

Our analysis expressed that if the effect of rotation parameter is
viewed, the action of the tunneling boson particles on the event horizon
will be different from the real particle's action and the calculated tunneling
probabilities are not just depending on the boson particles's energy $E$, angular momentum $j$
and particle's charge $e$ but it is also depending upon
the gauge potential $A_{\mu}$ of BH, rotation parameter $a$, curvature radius $\ell$ and BH
event horizon $r_+$. The corresponding Hawking temperature depending upon
the charge $q$, rotation parameter, curvature radius and event horizon of BHs. Moreover, the
Hawking temperature of BH increases due to the rotation parameter.
Furthermore, our analysis is similar to the analysis for charged scalar and fermion
particles tunneling from these BHs \cite{Gecim:2018sji,Meitei:2018mgo,Chen:2014eka,Zeng:2009zzd,Modak:2008tg}. The charged scalar and fermion particles
radiate through the BHs horizons with the same energy. Thus, the result shows that the electrically charged BTZ BH and BZT-like magnetic BH have slightly different emission spectrum and also Hawking temperature  due to different kind of charges such as electric charge or magnetic charge \cite{Li:2008ws,Li:2007zzg,Jiang:2007pn,He:2007zz,He:2007zzi,Li:2006rg}.
The Hawking temperature for magnetic BTZ like BH in Eq. (\ref{aTH}) can be calculated as follows
\begin{equation}\label{aTH}
\tilde{T}_{H}=\frac{r_{+}^{2}-4q^{2}\gamma^{2}l^{4}}{2\pi l^{2}r_+}.
\end{equation}
Moreover, the Hawking temperature for electrically charged BTZ BH is defined as follows
\begin{equation} 
\tilde{T}_{H}=\frac{2r_{+}^{3}-l^{4}Q^2}{4\pi l^{2}r_+^2}.
\end{equation}
We can observe that there is very slight difference in both temperatures. The Hawking temperature of BTZ like BH carries one extra arbitrary parameter $\gamma^{2}$ distigunish from electric BTZ BH.
\acknowledgements
A. \"{O}.~acknowledges
financial support provided under the Chilean FONDECYT Grant No. 3170035.

\end{document}